\documentclass[letter]{aa}
\usepackage[varg]{txfonts}
\usepackage{multirow}
\usepackage{natbib}
\usepackage{graphicx}

\begin{document}

\title{Tidal disruption event associated with the quasi-periodic eruptions from GSN 069: Possible disruption of a common envelope}

\author{Di Wang}
\institute{Department of Physics, Huazhong University of Science and Technology, Wuhan 430074, China}
\abstract
{X-ray quasi-periodic eruptions (QPEs) from the galactic nucleus have been found in several galaxies. Among them, GSN 069 is the only one with a tidal disruption event (TDE), which was recently found to have brightened again 9 years after the main outburst.}
{However, the origin of this TDE is still unclear. This Letter explores a particular model for the TDE.}
{By comparing the fallback time with observations, we found the TDE could not be caused by the disruption of the envelope of a single star in the tidal stripping model. Thus, we suggest that it is a disruption of a common envelope (CE).}
{By calculating the fallback rate of such a model, we reproduced the second peak in the observed TDE light curve. If this model is correct, this TDE will be the closest observation to a direct observation of CE, which has never been directly observed.
}
{}
\titlerunning{Disruption of a common envelope}
\maketitle
\section{Introduction}
GSN 069 is the first galaxy found to have X-ray quasi-periodic eruptions (QPEs), which originate in the region around the supermassive black hole (SMBH) \citep{miniutti2019nine}. The nature of the QPEs is still unclear, and its period $\sim$ 9 hrs may correspond to the orbital period of a body orbiting the SMBH, but the radiative processes and the nature of the body are controversial \citep{king2020gsn,king2022quasi,chen2022milli,zhao2022quasi,wang2022model,krolik2022quasiperiodic,lu2023quasi,xian2021x,sukova2021stellar,linial2023unstable,linial2023emri+}.

Although QPEs have since been seen in other galaxies \citep{giustini2020x,song2020possible,arcodia2021x,chakraborty2021possible}, GSN 069 remains the most exceptional one. It is the only galaxy with QPEs associated with a TDE, which has a $500\sim 800$ day timescale and is found to be re-brightening after 9 years since the main peak appeared \citep{shu2018long,miniutti2023repeating}. The abnormal carbon and nitrogen abundance ratio in its spectrum suggests that it is likely the disruption of a red giant \citep{sheng2021evidence}, whose core will continue to exist as a white dwarf (WD). This is consistent with the suggestion by \cite{king2020gsn} that the QPEs of GSN 096 come from a tidally stripped WD by the SMBH.

The origin of this TDE is puzzling. The orbital period of the disrupted star must be longer than the timescale of the TDE, which is much longer than the period of the QPEs. So if we assume that the disrupted star is the progenitor of the orbiting current star, then this star needs to have undergone a drastic change in orbital period after being partially disrupted. This change can be explained by the Hills mechanism, where the current orbiting star is captured by the SMBH from a binary system \citep{hills1988hyper,wang2022model}. Although the orbital period of the binary system and the SMBH can be very long at this time, the orbital velocity of the disrupted star with respect to the centre of mass of the binary also affects the timescale of the TDE.

In this Letter, we show that the disruption of the envelope of a single star always predicts a much shorter timescale than the observational value, which is shown in Section \ref{sec2}. So we suggest the TDE from GSN 069 is the disruption of a common envelope (CE), and we explain how we calculated its fallback rate in Section \ref{sec3}. We provide information on how we applied this model on GSN 069 in Section \ref{sec4} and discussions in Section \ref{sec5}.

\section{TDE of the envelope of a single star}\label{sec2}
The fallback timescale of TDE is determined by the specific binding energy of the most bound material
\begin{equation}
    \label{eq: timescale_tde sec2}
    t_{fb}=2\pi G M_{\mathrm{BH}}(-2\epsilon)^{-3/2},
\end{equation}
where the $G$ is the gravitational constant, $M_{\mathrm{BH}}$ is the mass of the SMBH, and $\epsilon$ is the specific binding energy.

If the orbit of the disrupted star is eccentric, the specific binding energy of the disrupted star $\epsilon_i$ is not equal to zero, and therefore it needs to be considered. The specific binding of the most bound debris is
\begin{equation}
    \label{eq: epsilon_sec2}
    \epsilon\approx \epsilon_i+\frac{G M_{\mathrm{BH}}}{R_t^2}R_*.
\end{equation}
Here the second term is the specific binding energy for the parabolic orbit \citep{gezari2021tidal}, $R_t$ is the disruption radius, and $R_*$ is the radius of the disrupted star.

We assume that the averaged period of $~9$ hrs for QPEs in GSN 069 corresponds to the orbital period of a WD orbiting an SMBH. If the TDE from GSN 069 is the disruption of an evolved star after it is captured, $\epsilon_i$ is equal to its specific binding energy. Then the fallback timescale of the TDE is smaller than the orbital period of the evolved star, which is also equal to the orbital period of the orbiting remnant WD. It is much smaller than the observed timescale of the TDE in GSN 069, which is about $500\sim 800$ days \citep{miniutti2023repeating}. In this case, the first term in Eq. \ref{eq: epsilon_sec2} is dominant, so a situation needs to be found to make it high enough to achieve the observed timescale.

So the disruption must occur before the capture of the WD. In this case, the velocity of the disrupted envelope contains two components: the orbital velocity of the disrupted star $v_*$ and the velocity of the binary $v_b$ relative to the SMBH. Assuming the binary is in a parabolic orbit, the specific binding energy of the disrupted star is 
\begin{equation}
    \label{eq: epsilon0_sec2}
    \epsilon_i\approx v_*v_b\cos\theta,
\end{equation}
where $\theta$ is the angle between $v_*$ and $v_b$. The $v_*^2/2$ term is ignored. 

Assuming the WD is captured at the periastron by the Hills mechanism, the capture of the WD corresponds to $\theta= \pi$, so the timescale corresponding to the specific binding energy $\epsilon_i\approx-v_*v_b$ is the orbital period of captured WD. If $\theta$ is close to $0$ while the disruption occurs, corresponding to the situation that the disrupted star is the companion of the captured WD, the disrupted envelope is subsequently fully unbound. In this case, Eq. \ref{eq: epsilon_sec2} leads to a positive binding energy because $\epsilon_i=v_*v_b$ is much higher than the second term. So to produce the observational long timescale, $\epsilon_i$ needs to be close to zero, which means $\theta$ needs to be close to $\pi/2$ at the moment of disruption.

So the disruption distance needs to be sufficiently larger than the capture distance to ensure the angle $\theta$ is close to $\pi/2$. The capture distance is \citep{amaro2018relativistic}
\begin{equation}
    R_{H}=\left(\frac{M_{\mathrm{BH}}}{m_b}\right)^{1/3}a_b.
\end{equation}
Here $m_*$ and $m_c$ are the mass of the disrupted star and its companion, respectively. We note that $m_b=m_*+m_c$ is the total mass of the binary, and $a_b$ is the separation of the binary. 

\begin{figure}
    \centering
    \includegraphics[width=0.5\textwidth]{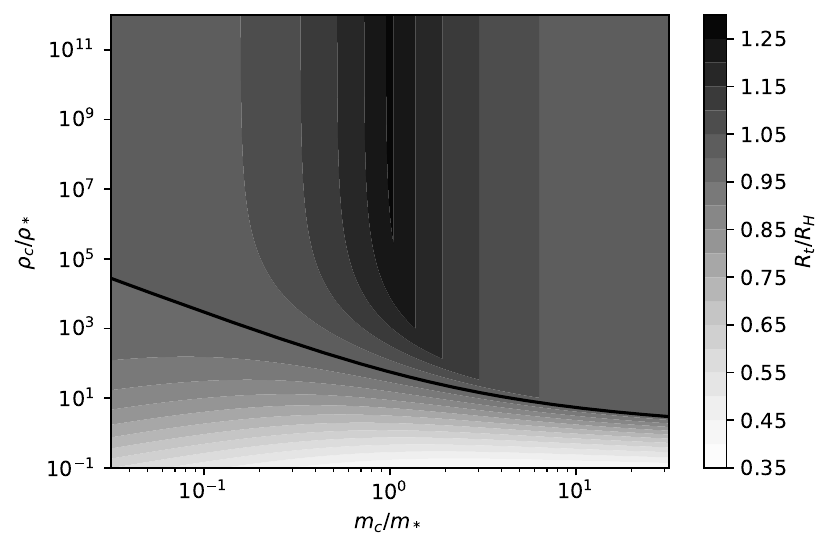}
    \caption{Value of the upper limit of $R_t/R_H$ for the companion star of various densities and masses. Here $\rho_c$ and $\rho_*$ are the average density of the companion and the disrupted star, respectively. The solid line represents $R_t/R_H=1$.\label{fig: hill}}
\end{figure}

The disruption radius is \citep{gezari2021tidal}
\begin{equation}
    R_t=\left(\frac{M_{\mathrm{BH}}}{m_*}\right)^{1/3}R_*.
\end{equation}
So the ratio of these two radii is 
\begin{equation}
    \frac{R_t}{R_H}=\left(\frac{m_*+m_c}{m_*}\right)^{1/3}\frac{R_*}{a_b},
\end{equation}
which is independent of the mass of the SMBH. To avoid the disruption of the evolved star by the companion, the separation needs to satisfy
\begin{equation}
    a_b>\left(\frac{M_{c}}{m_*}\right)^{1/3}R_*.
\end{equation}
And the companion is outside the envelope of the evolved star, requiring 
\begin{equation}
    a_b>R_*+R_c.
\end{equation}
The above two conditions together give the upper limit of $R_t/R_H$. As shown in Fig. \ref{fig: hill}, $R_t$ can be larger than $R_H$ when the companion is much more compact than the disrupted star, and the maximum $R_t$ is about $1.25R_H$.

The time for the binary to move from distance $R_t$ to periastron $R_H$ is 
\begin{equation}\label{eq: timescale of motion}
    t=-\frac{\frac{1}{2}\tan\frac{f}{2}+\frac{1}{6}\tan^3\frac{f}{2}}{\bar{n}},
\end{equation}
where $f$ is the true anomaly, and it can be obtained by $\sec^2\frac{f}{2}=R_t/R_H$ ($f$ is negative). Furthermore, $\bar{n}$ is the averaged angular velocity 
\begin{equation}
    \bar{n}=\sqrt{\frac{G M_{\mathrm{BH}}}{8R_H^3}}=\sqrt{\frac{G (m_*+m_c)}{8a_b^3}}=\frac{n_b}{2\sqrt{2}}.
\end{equation}
Here $n_b=\sqrt{G (m_*+m_c)/a_b^3}$ is the angular velocity of the binary. So the change in the direction of the orbital position caused by orbital motion in the binary is 
\begin{equation}\label{eq: angle of orbital motion}
    \Delta\eta=-n_b t=\sqrt{2}\tan\frac{f}{2}+\frac{\sqrt{2}}{3}\tan^3\frac{f}{2}.
\end{equation}
Fig. \ref{fig: hill} gives an upper limit of $R_t/R_H$ of about 1.25, which corresponds to a minimum $f$. So the maximum $\Delta\eta_{max}$ can be obtained by Eq. \ref{eq: angle of orbital motion} and it is about 0.766 rad. Due to the maximum of $R_t/R_H$ being small, the change in the direction of the orbital velocity of the binary in the parabolic orbit can be ignored. So the change of $\theta$ in Eq. \ref{eq: epsilon0_sec2} is roughly equal to $\Delta\eta$. Due to $\theta=\pi$ at the capture distance, the $\theta$ at the disruption distance is $\pi-\Delta \theta$.

So $\Delta\theta_{max}$ corresponds to a maximum $\epsilon_i \approx -0.72 v_*v_b$. Combined with $t_{fb}\sim (-\epsilon)^{-3/2}$, the maximum timescale of TDE in this situation is about 1.6 times the orbital period of the captured WD $\sim 14.4$ hrs, which is still much smaller than the observational value.

Now, we can safely rule out the situations in which TDE from GSN 069 is the disruption of the envelope of either of the stars in the binary. The one remaining possible scenario is that it is the disruption of a CE that contains the double stars in it. In this scenario, the velocity of the envelope is the orbital velocity, and $\epsilon_i$ is zero for a parabolic orbit. So the observational timescale of the TDE can be produced easily.

\section{TDE of the CE} \label{sec3}
The structure of the CE is still unclear \citep{ivanova2013common}. Here we only consider the effect of the differential rotation of the CE. We adopted a similar model as early works\footnote{The previous works used a 1D model in the spherical coordinate, where the rotational velocity only depends on the distance to the centre. The following calculations need the full structure of the CE. Thus, we adopted the cylindrical coordinate and assume that the rotational velocity only depends on the distance to the rotational axis.} \citep{meyer1979formation,meng2017common,song2020structure}. The envelope is co-rotating with the binary at the centre and the rotational velocity decreases outwards, following a radial power law. The rotational angular velocity can be described by 
\begin{equation}
    \Omega= \begin{cases}\Omega_0, & s \leq s_0 \\ \Omega_0\left(\frac{s}{s_0}\right)^{-\alpha}, & s>s_0\end{cases}
,\end{equation}
where $s$ is the distance to the rotational axis, and $s_0$ and $\Omega_0$ are the distance and the rotational angular velocity at the boundary of the binary, respectively.

\subsection{Effect of rotation on specific binding energy}
Under the frozen-in approximation, the fallback rate of the TDE is \citep{gezari2021tidal} 
\begin{equation}\label{eq: fallback rate}
    \dot{M}=\frac{d M}{d \epsilon} \frac{d \epsilon}{d t}=\frac{2 \pi}{3}\left(G M_{\mathrm{BH}}\right)^{2 / 3} \frac{d M}{d \epsilon} t^{-5 / 3}.
\end{equation}
It strongly depends on the distribution of the specific binding energy in the star. 

The rotation can affect the specific binding energy by providing an additional speed. Assuming the centre of mass (COM) is on a parabolic orbit, the specific binding energy with rotation is \citep{golightly2019tidal}
\begin{equation}\label{eq: specific binding energy}
    \epsilon\approx \left(\vec{v} \times \vec{\Omega}+\frac{G M_{\mathrm{BH}}}{R_t^3} \vec{R}_t\right) \cdot \vec{r}.
\end{equation}
Here $\vec{v}$ and $\vec{R_t}$ are the velocity and the position of the COM relative to the central black hole while the star begins to be disrupted, respectively. Furthermore, $\vec{\Omega}$ is the rotational angular velocity, and $\vec{r}$ is the position within the star relative to the COM. Only the lowest order of the rotation has been considered for this Letter.

The specific binding energy can also be viewed as a projection of the position $\vec{r}$ in the direction $\vec{R}_{P}=\vec{v} \times \vec{\Omega}+\frac{G M_{\mathrm{BH}}}{R_t^3} \vec{R}_t$, and it only depends on the projection distance. For the case of the rigid rotation, the projection vector $\vec{R}_{P}$ is independent of the position $\vec{r}$, as in the case of no rotation ($\vec{\Omega}=0$). Thus, the star with rigid rotation has a similar distribution to the specific binding energy with the star without rotation. Both of them have isoenergetic surfaces that are the plane perpendicular to the projection vector $\vec{R}_{P}$, and only the magnitude of the specific binding energy is different. So the disruption of the star with rigid rotation has a similar evolution to the fallback rate with the star without rotation, as shown in \cite{golightly2019tidal}.

If the rotation is differential, the projection vector $\vec{R}_{P}$ is dependent on the position $\vec{r}$. Then the equipotential surfaces are no longer planar but curved, which depends on the distribution of the rotation $\vec{\Omega}$. 

Here we consider a simple case. A CE with mass $M_*$ and radius $R_*$ is disrupted at the periastron, and vector $\vec{\Omega}$ is perpendicular to the orbital plane, which means that the $\hat{v}$, $\hat{\Omega}$, and $\hat{R}_t$ are perpendicular to each other. With $\hat{\Omega}$ as the z-axis and $\hat{R}_t$ as the x-axis, using the cylindrical coordinate $(s, \phi, z)$, equation \ref{eq: specific binding energy} can be rewritten as
\begin{equation}\label{eq: specific binding energy 2}
    \begin{aligned}
    \epsilon&=s\left(\frac{GM_{\mathrm{BH}}}{R_t^2}+\Omega v\right)\cos\phi\\
    &=\epsilon_0 \bar{s}(1+k \bar{s}^{-\alpha})\cos\phi.
    \end{aligned}
\end{equation}
Here $\bar{s}\equiv s/R_*$ is the normalized distance. Furthermore, $\epsilon_0\equiv  GM_{\mathrm{BH}}R_*/R_t^2 $ is the standard specific binding energy of the envelope without rotation. We assume that the angular velocity is the power law, for example $\Omega=\Omega_{s}\bar{s}^{-\alpha}$, where $\Omega_{s}$ is the rotational velocity on the surface of the envelope. The parameter $k$ is
\begin{equation}\label{eq: k}
    \begin{aligned}
    k&\equiv \frac{\Omega_s v}{(GM_{\mathrm{BH}}/R_t^2)}=\Omega_s\sqrt{\frac{2 R_t^3}{G M_{\mathrm{BH}}}}\\
    &=\sqrt{2}\frac{\Omega_s}{\Omega_*}=k_0\bar{s}_0^{\alpha-\frac{3}{2}},
    \end{aligned}
\end{equation}
where $k_0\equiv\sqrt{2M_c/M_*}$. In the derivation of the above equation, the velocity of the parabolic orbit $v=\sqrt{2GM_{\mathrm{BH}}/R_t}$, and the disrupted radius $R_t=(M_{\mathrm{BH}}/M_*)^{1/3}R_*$ is adopted. In addition, $\Omega_*\equiv \sqrt{GM_*/R_*^3}$ is the break-up rotation velocity of the star.

\begin{figure}
    \centering
    \includegraphics[width=0.5\textwidth]{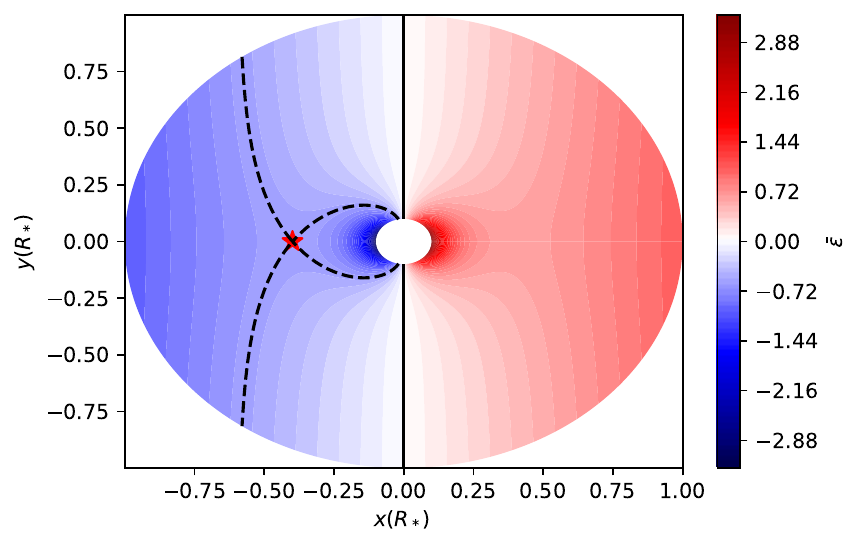}
    \caption{Distribution of $\bar{\epsilon}$ with $k_0=1$, $\bar{s}_0=0.1$, and $\alpha=3$. The red star is the stationary point, and the corresponding $\bar{\epsilon}$ is represented by the dashed line. The solid line represents $\bar{\epsilon}=0$, and $\bar{\epsilon}$ is negative on its left side and positive on its right side.\label{fig: epsilon1}}
\end{figure}

We show an example of the  $\bar{\epsilon}=\epsilon/\epsilon_0$ distribution  with $k_0=1$, $\bar{s}=0.1$, and $\alpha=3$ in Fig. \ref{fig: epsilon1}, and the regions of interest are with $\bar{\epsilon}<0$. There is a stationary point near which the change in $\bar{\epsilon}$ slows down. The isoenergetic surface through this point delineates the distribution of $\bar{\epsilon}$ into four regions. As shown in Fig. \ref{fig: epsilon1}, the region with the lowest $\bar{\epsilon}$ is to the right of the stationary point, the left side region has a higher $\bar{\epsilon}$, and the highest $\bar{\epsilon}$ is distributed in both the upper and lower regions. A smaller gradient of $\bar{\epsilon}$ near the stationary point predicts a higher $\mathrm{d}M/\mathrm{d}\epsilon$ here and therefore a peak in the fallback rate.

Because the SMBH is in the direction of $\phi=\pi$, the position of the stationary point can be obtained by solving for the minimum value of equation \ref{eq: specific binding energy 2} with $\cos\phi=-1$,
\begin{equation}\label{eq: s_p}
    \begin{aligned}
    \bar{s}_{p2}&=(k(\alpha-1))^{1/\alpha},
    \end{aligned}
\end{equation}
which corresponds to the specific binding energy
\begin{equation}\label{eq: epsilon_p}
    \bar{\epsilon}_{p2}=-\frac{\alpha(k(\alpha-1))^{1/\alpha}}{\alpha-1}.
\end{equation}
The subscript p2 indicates that it is the second peak. Then the time of the second peak can be obtained by Eq. \ref{eq: timescale_tde sec2}.

The apparent time of the second peak depends on the distribution of the angular velocity in the envelope. For a given $\alpha$, the higher $k$ is, the smaller $\bar{\epsilon}$, which predicts the earlier second peak. 

\subsection{Fallback rate}
\begin{figure}[h]
    \centering
    \includegraphics[width=0.5\textwidth]{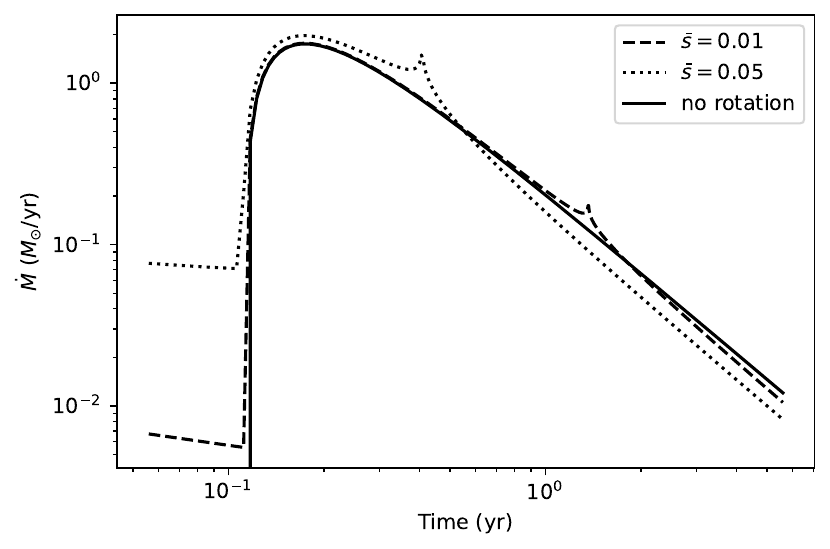}
    \caption{Fallback rate for the different rotation. Comparing the case of no rotation, the disruption of the envelope with differential rotation shows a second peak in the decay phase. The larger the core, the earlier the peak appears. The earlier fallback with differential rotation is caused by the innermost material, which has the lowest specific binding energy. \label{fig: lc}}
\end{figure}

The mass of disrupted star can be expressed as $M=\iiint \rho \bar{s}\,\mathrm{d}\bar{s}\mathrm{d}\bar{z}\mathrm{d}\phi$. By transformation from cylindrical coordinate $(\bar{s}, \phi, \bar{z})$ to coordinate $(\bar{s}, \bar{\epsilon}, \bar{z})$, it becomes
\begin{equation}
    \begin{aligned}
    \frac{\mathrm{d}M}{\mathrm{d}\epsilon}&=\frac{-2}{\epsilon_0}\iint \left|\frac{\partial(\bar{s},\phi,\bar{z})}{\partial(\bar{s},\bar{\epsilon},\bar{z})}\right|\rho \bar{s} \,\mathrm{d}\bar{s}\mathrm{d}\bar{z}\\
    &=\frac{2}{\epsilon_0}\int\frac{\rho_z\bar{s} }{\sqrt{\bar{s}^2(1+k\bar{s}^{-\alpha})^2-\bar{\epsilon}^2}} \,\mathrm{d}\bar{s}.
    \end{aligned}
\end{equation}
Here $\rho_z\equiv \int\rho\mathrm{d}z$ is the column density. Then the fallback rate can be obtained by Eq. \ref{eq: fallback rate}. Here we show the fallback rate with $R_*=1\,R_{\odot}$, $M_*=1\,M_{\odot}$, $M_{\mathrm{BH}}=10^6 \,M_{\odot}$, $\alpha=3$, and $k_0=1$  in Fig. \ref{fig: lc}. The disrupted star is assumed to be spherical, and the density is assumed to be constant.

Comparing the curve of the envelope without rotation, the disruption of the envelope with differential rotation as expected shows a second peak. A larger core leads to the stationary point further away from the core, thus producing earlier peaks that are also stronger. Before this peak appears, the curve decays similarly to the case without rotation, due to the small rotation of the material in the outer layers of the disrupted envelope. The fallback before the main peak is the accretion of the innermost material, which has the smallest binding energy.

\section{Application to the TDE in GSN 069} \label{sec4}
Although our calculated fallback rate reproduces the second peak of the observational light curve, a direct comparison between the full light curve and our model is inappropriate. Under the frozen-in approximation, pressure and self-gravity in the accretion flow are neglected. In Fig. \ref{fig: lc} it can be seen that the material near the core has the smallest fallback timescale, and it may interact with material near the stationary point thereby exchanging energy while moving outwards, which leads to the broadening of the second peak. Furthermore, since the stationary point corresponds to a region of high density in the accretion flow, self-gravity could potentially lead to self-bound material around the stationary point and thus narrow the second peak. However, these effects only affect the shape of the second peak and not the time of its appearance.

The gravitation of the core significantly affects the fallback curve and can make it go from a -5/3 decay to a -9/4 decay, which is the curve of a partial tidal disruption event \citep{coughlin2019partial}. Considering that material closer to the core is affected by a stronger gravitation of the core relative to the gravitation of the SMBH, the fallback timescales of material near the stationary point are expected to be modified more than those of the most bounded material. Thus, the ratio of the time of appearance of the second peak to the main peak should become larger. On the other hand, the captured star may also affect the motion of the disrupted envelope. These effects cannot be solved analytically, so we ignore them for now in the discussion that follows.

The density distribution of the disrupted envelope also significantly affects the fallback curve. The realistic density distribution of the CE should be quite different from the constant density we assumed. Although its exact distribution is not known, the stationary point that appears in the specific binding energy distribution still produces a peak in the fallback rate curve as long as the density distribution is relatively smooth.

Therefore, the only parameter that can be used to make comparisons with observations is the relative time between the appearance of the second peak and the main peak. The specific binding energy of the most bound material is $\bar{\epsilon}_{p1}=-(1+k)$ by Eq. \ref{eq: specific binding energy 2}, where the subscript p1 represents the main peak. The ratio of the fallback time between the main peak and the second peak is
\begin{equation}
    \frac{t_{p2}}{t_{p1}}=\left(\frac{\bar{\epsilon}_{p2}}{\bar{\epsilon}_{p1}}\right)^{-3/2}=\left(\frac{\alpha(k(\alpha-1))^{1/\alpha}}{(\alpha-1)(1+k)}\right)^{-3/2},
\end{equation}
which only depends on the rotation distribution of the CE.

\cite{miniutti2023repeating} fitted the light curve of the two peaks of the TDE from GSN 069 using a formula of Gaussian rise plus power-law decay with the index -9/4. They found $t^{obs}_{p1} \simeq 500-800 \mathrm{~d}$ and $t^{obs}_{p2} \simeq 3750-3900 \mathrm{~d}$. We adopted their median value and obtained $(t_{p2}/t_{p1})_{obs}\simeq 4.7-7.8$. 

They found the accreted mass is about $M_{\mathrm{accr}}=0.23 M_{\odot}$ by integrating the light curve as well. So the mass of the disrupted envelope is about $0.46 M_{\odot}$ considering the unbound material. The mass of the WD is about $0.21 M_{\odot}$ \citep{king2020gsn}. Assuming the companion has a similar mass to the WD, we have $k_0=1$.

\begin{figure}
    \centering
    \includegraphics[width=0.5\textwidth]{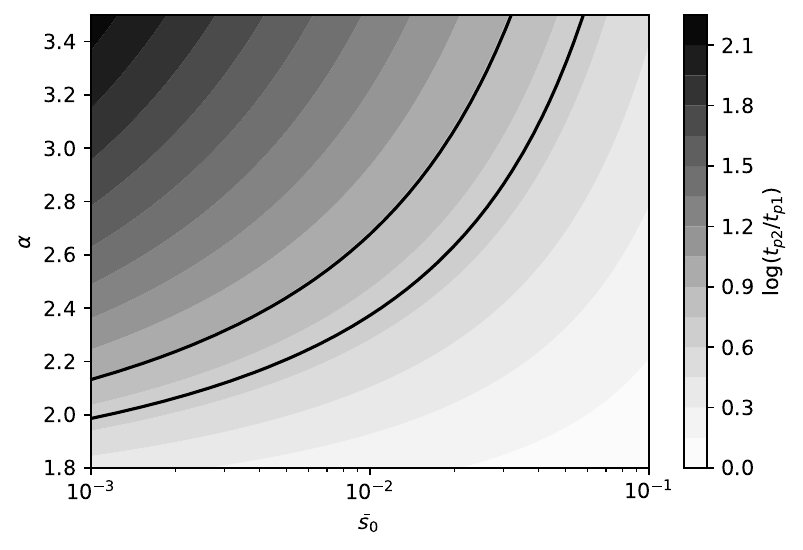}
    \caption{Constraint on the rotation decay index $\alpha$ and the relative scale of the core $\bar{s}_0$. The two solid lines represent $(t_{p2}/t_{p1})_{obs}=$ 4.7 and 7.8. The area they enclose is the constrained region of the parameters.\label{fig: app1}}
\end{figure}

Now the rotation distribution can be constrained. As shown in Fig. \ref{fig: app1}, the rotation decay index $\alpha$ and the relative scale of core $\bar{s}_0$ is constrained in a narrow region. If $\alpha$ is considered to be three as suggested by \cite{meyer1979formation}, $\bar{s}_0$ is about 0.02. The radius of the envelope can be obtained by the fallback time $R_* \simeq 5-12 R_{\odot}$ \citep{miniutti2023repeating}. Thus, the separation of the binary in the core is about $11-26 R_{\oplus}$. \cite{wang2022model} adopted the separation about $8.3 \sim 83 R_{\oplus}$ to estimate the eccentricity $0.97\sim 1$ by the Hills mechanism.

As per the discussions in section \ref{sec2}, the TDE occurs before the capture of the WD at the periastron. So $\hat{v}$ is no longer perpendicular to $\hat{R}_t$ when the envelope is disrupted. However, it only slightly affects the constraint on the rotation disruption. Detailed calculations are shown in Appendix \ref{appendix}.

\section{Discussion} \label{sec5}
While the tidal disruption of a CE can explain the second peak of the light curve of GSN 069, there is also a serious issue that the lifetime of a CE is very short, and the probability that we will see such events directly is very low. Simulations show that the maximum lifetime is only about $10^5$ times the orbital period of the binary in the core \citep{gagnier2023post}. This timescale is only 273 years for a 1-day orbital period. 

Another possibility is that the disrupted envelope is a circumbinary disk, which might be formed by the remaining material in the post-CE stage \citep{ivanova2013common}. If so, the circumbinary disk must not be a Keplerian disk, because it would be disrupted by a very slight tidal force.

To understand the detailed nature of the disrupted envelope, hydrodynamics simulations are necessary. Not only do the fallback rate curves need to take into account the interactions between the accreted material and the effect of the gravitation by the core and the captured WD, but also the realistic density and rotation distributions of the CE need to be simulated. If it can be confirmed that the TDE from GSN 069 is a disruption of the CE, then it is the closest one can get to a direct observation of the CE.

Our model also strongly relies on the validity of the binary model for QPEs following the capture of the star through the Hills mechanism. However, there are some alternative models for QPEs \citep{sniegowska2020possible,ingram2021self,raj2021disk,pan2022disk,kaur2023magnetically}. Future improvements in the understanding of the QPEs are very important for checking our model.

On the other hand, some other TDEs have a re-brightening peak \citep{jiang2019infrared,wevers2023live,malyali2023rebrightening}. Our model of the disruption of a differentially rotating envelope provides an alternative interpretation for such TDEs, even if they are not associated with QPEs. Not only does the envelope of the CE have differential rotation, but the envelope of the red giant may have it as well \citep{klion2017diagnostic}. Neither case necessarily requires the presence of QPEs. Considering the short lifetime of the CE, These sources are a little more likely to be the disruption of a red giant. Furthermore, our model can also produce a third peak if the disrupted envelope is in an eccentric orbit, which will make the binding energy of the stationary point further away from SMBH positive.

We cannot currently conclude whether these re-brightening TDEs are a disruption to the differentially rotating envelope because, as discussed in Section \ref{sec4}, only the time of the appearance of the second peak relative to the main peak is available for comparison with observations. Models that take into account density distributions, self-gravity, and gravity by the core may be able to provide more accurate fallback rate profiles that can be used to determine the nature of these re-brightening TDEs, which is beyond the scope of this Letter. Therefore, the study of more detailed models and numerical simulations is necessary in the future.

\begin{acknowledgements}
    This work was supported by the National Key Research and Development Program of China (No. 2020YFC2201400) and National SKA Program of China  (2020SKA0120300).
\end{acknowledgements}

\bibliography{ref} 

\begin{thebibliography}{38}
\expandafter\ifx\csname natexlab\endcsname\relax\def\natexlab#1{#1}\fi

\bibitem[{Amaro-Seoane(2018)}]{amaro2018relativistic}
Amaro-Seoane, P. 2018, Living Reviews in Relativity, 21, 4

\bibitem[{Arcodia {et~al.}(2021)Arcodia, Merloni, Nandra, Buchner, Salvato,
  Pasham, Remillard, Comparat, Lamer, Ponti, {et~al.}}]{arcodia2021x}
Arcodia, R., Merloni, A., Nandra, K., {et~al.} 2021, Nature, 592, 704

\bibitem[{Chakraborty {et~al.}(2021)Chakraborty, Kara, Masterson, Giustini,
  Miniutti, \& Saxton}]{chakraborty2021possible}
Chakraborty, J., Kara, E., Masterson, M., {et~al.} 2021, The Astrophysical
  Journal Letters, 921, L40

\bibitem[{Chen {et~al.}(2022)Chen, Qiu, Li, \& Liu}]{chen2022milli}
Chen, X., Qiu, Y., Li, S., \& Liu, F. 2022, The Astrophysical Journal, 930, 122

\bibitem[{Coughlin \& Nixon(2019)}]{coughlin2019partial}
Coughlin, E.~R. \& Nixon, C. 2019, The Astrophysical Journal Letters, 883, L17

\bibitem[{Gagnier \& Pejcha(2023)}]{gagnier2023post}
Gagnier, D. \& Pejcha, O. 2023, Astronomy \& Astrophysics, 674, A121

\bibitem[{Gezari(2021)}]{gezari2021tidal}
Gezari, S. 2021, Annual Review of Astronomy and Astrophysics, 59, 21

\bibitem[{Giustini {et~al.}(2020)Giustini, Miniutti, \& Saxton}]{giustini2020x}
Giustini, M., Miniutti, G., \& Saxton, R.~D. 2020, Astronomy \& Astrophysics,
  636, L2

\bibitem[{Golightly {et~al.}(2019)Golightly, Coughlin, \&
  Nixon}]{golightly2019tidal}
Golightly, E.~C., Coughlin, E.~R., \& Nixon, C. 2019, The Astrophysical
  Journal, 872, 163

\bibitem[{Hills(1988)}]{hills1988hyper}
Hills, J.~G. 1988, Nature, 331, 687

\bibitem[{Ingram {et~al.}(2021)Ingram, Motta, Aigrain, \&
  Karastergiou}]{ingram2021self}
Ingram, A., Motta, S.~E., Aigrain, S., \& Karastergiou, A. 2021, Monthly
  Notices of the Royal Astronomical Society, 503, 1703

\bibitem[{Ivanova {et~al.}(2013)Ivanova, Justham, Chen, De~Marco, Fryer,
  Gaburov, Ge, Glebbeek, Han, Li, {et~al.}}]{ivanova2013common}
Ivanova, N., Justham, S., Chen, X., {et~al.} 2013, The Astronomy and
  Astrophysics Review, 21, 1

\bibitem[{Jiang {et~al.}(2019)Jiang, Wang, Mou, Liu, Dou, Sheng, \&
  Wang}]{jiang2019infrared}
Jiang, N., Wang, T., Mou, G., {et~al.} 2019, The Astrophysical Journal, 871, 15

\bibitem[{Kaur {et~al.}(2023)Kaur, Stone, \& Gilbaum}]{kaur2023magnetically}
Kaur, K., Stone, N.~C., \& Gilbaum, S. 2023, Monthly Notices of the Royal
  Astronomical Society, stad1894

\bibitem[{King(2020)}]{king2020gsn}
King, A. 2020, Monthly Notices of the Royal Astronomical Society: Letters, 493,
  L120

\bibitem[{King(2022)}]{king2022quasi}
King, A. 2022, Monthly Notices of the Royal Astronomical Society, 515, 4344

\bibitem[{Klion \& Quataert(2017)}]{klion2017diagnostic}
Klion, H. \& Quataert, E. 2017, Monthly Notices of the Royal Astronomical
  Society: Letters, 464, L16

\bibitem[{Krolik \& Linial(2022)}]{krolik2022quasiperiodic}
Krolik, J.~H. \& Linial, I. 2022, The Astrophysical Journal, 941, 24

\bibitem[{Linial \& Metzger(2023)}]{linial2023emri+}
Linial, I. \& Metzger, B.~D. 2023, arXiv preprint arXiv:2303.16231

\bibitem[{Linial \& Sari(2023)}]{linial2023unstable}
Linial, I. \& Sari, R. 2023, The Astrophysical Journal, 945, 86

\bibitem[{Lu \& Quataert(2023)}]{lu2023quasi}
Lu, W. \& Quataert, E. 2023, Monthly Notices of the Royal Astronomical Society,
  524, 6247

\bibitem[{Malyali {et~al.}(2023)Malyali, Liu, Rau, Grotova, Merloni, Goodwin,
  Anderson, Miller-Jones, Kawka, Arcodia, {et~al.}}]{malyali2023rebrightening}
Malyali, A., Liu, Z., Rau, A., {et~al.} 2023, Monthly Notices of the Royal
  Astronomical Society, 520, 3549

\bibitem[{Meng \& Podsiadlowski(2017)}]{meng2017common}
Meng, X. \& Podsiadlowski, P. 2017, Monthly Notices of the Royal Astronomical
  Society, 469, 4763

\bibitem[{Meyer \& Meyer-Hofmeister(1979)}]{meyer1979formation}
Meyer, F. \& Meyer-Hofmeister, E. 1979, Astronomy and Astrophysics, vol. 78,
  no. 2, Sept. 1979, p. 167-176., 78, 167

\bibitem[{Miniutti {et~al.}(2023)Miniutti, Giustini, Arcodia, Saxton, Read,
  Bianchi, \& Alexander}]{miniutti2023repeating}
Miniutti, G., Giustini, M., Arcodia, R., {et~al.} 2023, Astronomy and
  Astrophysics, 670, A93

\bibitem[{Miniutti {et~al.}(2019)Miniutti, Saxton, Giustini, Alexander, Fender,
  Heywood, Monageng, Coriat, Tzioumis, Read, {et~al.}}]{miniutti2019nine}
Miniutti, G., Saxton, R., Giustini, M., {et~al.} 2019, Nature, 573, 381

\bibitem[{Pan {et~al.}(2022)Pan, Li, Cao, Miniutti, \& Gu}]{pan2022disk}
Pan, X., Li, S.-L., Cao, X., Miniutti, G., \& Gu, M. 2022, The Astrophysical
  Journal Letters, 928, L18

\bibitem[{Raj \& Nixon(2021)}]{raj2021disk}
Raj, A. \& Nixon, C. 2021, The Astrophysical Journal, 909, 82

\bibitem[{Sheng {et~al.}(2021)Sheng, Wang, Ferland, Shu, Yang, Jiang, \&
  Chen}]{sheng2021evidence}
Sheng, Z., Wang, T., Ferland, G., {et~al.} 2021, The Astrophysical Journal
  Letters, 920, L25

\bibitem[{Shu {et~al.}(2018)Shu, Wang, Dou, Jiang, Wang, \& Wang}]{shu2018long}
Shu, X., Wang, S., Dou, L., {et~al.} 2018, The Astrophysical Journal Letters,
  857, L16

\bibitem[{Sniegowska {et~al.}(2020)Sniegowska, Czerny, Bon, \&
  Bon}]{sniegowska2020possible}
Sniegowska, M., Czerny, B., Bon, E., \& Bon, N. 2020, Astronomy \&
  Astrophysics, 641, A167

\bibitem[{Song {et~al.}(2020{\natexlab{a}})Song, Shu, Sun, Xue, Jin, Zhang,
  Jiang, Dou, \& Wang}]{song2020possible}
Song, J., Shu, X., Sun, L., {et~al.} 2020{\natexlab{a}}, Astronomy \&
  Astrophysics, 644, L9

\bibitem[{Song {et~al.}(2020{\natexlab{b}})Song, Meng, Podsiadlowski, \&
  Cui}]{song2020structure}
Song, R., Meng, X., Podsiadlowski, P., \& Cui, Y. 2020{\natexlab{b}}, Astronomy
  \& Astrophysics, 633, A41

\bibitem[{Sukov{\'a} {et~al.}(2021)Sukov{\'a}, Zaja{\v{c}}ek, Witzany, \&
  Karas}]{sukova2021stellar}
Sukov{\'a}, P., Zaja{\v{c}}ek, M., Witzany, V., \& Karas, V. 2021, The
  Astrophysical Journal, 917, 43

\bibitem[{Wang {et~al.}(2022)Wang, Yin, Ma, \& Wu}]{wang2022model}
Wang, M., Yin, J., Ma, Y., \& Wu, Q. 2022, The Astrophysical Journal, 933, 225

\bibitem[{Wevers {et~al.}(2023)Wevers, Coughlin, Pasham, Guolo, Sun, Wen,
  Jonker, Zabludoff, Malyali, Arcodia, {et~al.}}]{wevers2023live}
Wevers, T., Coughlin, E., Pasham, D., {et~al.} 2023, The Astrophysical Journal
  Letters, 942, L33

\bibitem[{Xian {et~al.}(2021)Xian, Zhang, Dou, He, \& Shu}]{xian2021x}
Xian, J., Zhang, F., Dou, L., He, J., \& Shu, X. 2021, The Astrophysical
  Journal Letters, 921, L32

\bibitem[{Zhao {et~al.}(2022)Zhao, Wang, Zou, Wang, \& Dai}]{zhao2022quasi}
Zhao, Z., Wang, Y., Zou, Y., Wang, F., \& Dai, Z. 2022, Astronomy \&
  Astrophysics, 661, A55

\end{thebibliography}
\bibliographystyle{aa}

\begin{appendix}

\section{The disruption before reaching periastron}\label{appendix}
If the disruption occurs before the envelope reaches periastron, that is, $\beta\equiv R_t/R_p>1$, $\hat{v}$ is no longer perpendicular to $\hat{R}_t$ when the envelope is disrupted. Assuming it rotates in the direction of $\hat{R}_t$ by an angle $\gamma$, then Eq. \ref{eq: specific binding energy 2} becomes
\begin{equation}
    \epsilon=\epsilon_0\bar{s}(\cos\phi+k\bar{s}^{-\alpha}\cos(\phi+\gamma)).
\end{equation}
For a parabolic orbit, $\gamma$ can be obtained by 
\begin{equation}
    \sin \gamma=\frac{\sqrt{\beta}}{2}\sin f,
\end{equation}
where the true anomaly $f$ is obtained by $\cos f=2/\beta -1$. 

\begin{figure}[h]
    \centering
    \includegraphics[width=0.5\textwidth]{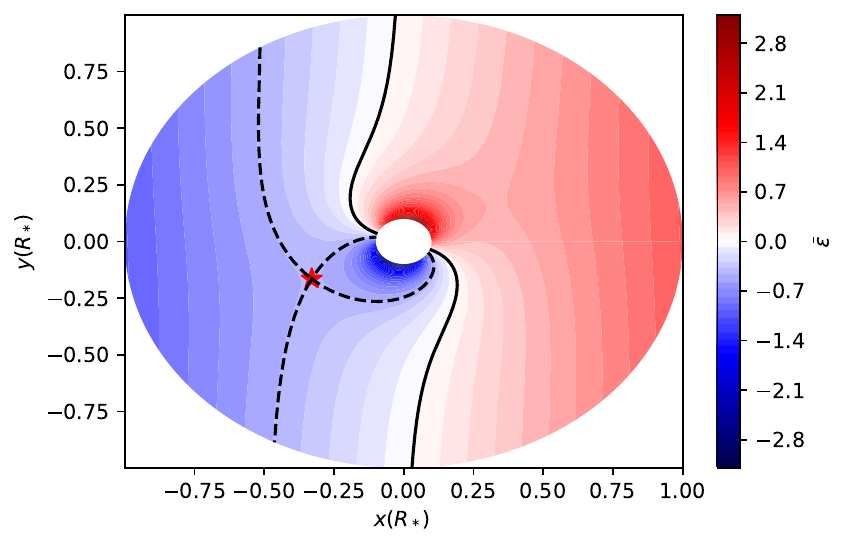}
    \caption{Same as Fig. \ref{fig: epsilon1}, but with $\beta=10$.\label{fig: epsilon2}}
\end{figure}

\begin{figure}[h]
    \centering
    \includegraphics[width=0.5\textwidth]{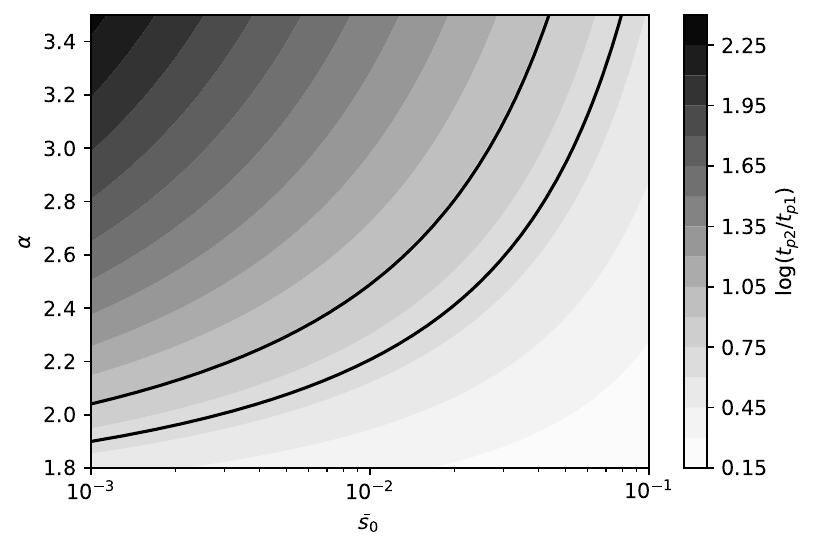}
    \caption{Same as Fig. \ref{fig: app1}, but with $\beta=10$.\label{fig: app2}}
\end{figure}

As shown in Fig. \ref{fig: epsilon2}, the $\bar{\epsilon}$ distribution with $\beta=10$ is rotated compared with the distribution with $\beta=1$. The position of the stationary point can be obtained by solving 

\begin{equation}
    \begin{aligned}
        \frac{\partial \bar{\epsilon}}{\partial \bar{s}}&=\bar{s}^{-\alpha}\left[(1-\alpha) k \cos (\phi+\gamma)+\bar{s}^\alpha \cos \phi\right]=0\\
        \frac{\partial \epsilon}{\bar{s} \partial \phi}&=-k \bar{s}^{-\alpha} \sin (\phi+\gamma)-\sin \phi=0.
    \end{aligned}
\end{equation}

Then the time of the second peak can be obtained by the corresponding $\epsilon$. As shown in the Fig. \ref{fig: app2}, the constraint on the rotation distribution with $\beta=10$ gives a slightly smaller $\alpha$ and a slightly larger $\bar{s}$ than those with $\beta=1$.

\end{appendix}

\end{document}